\documentclass[%
 reprint,
 amsmath,amssymb,
 aps,
]{revtex4-2}

\usepackage{graphicx}
\usepackage{dcolumn}
\usepackage{bm}

\begin{document}


\title{Strategies and safety simulations for ultrasonic cervical spinal cord neuromodulation}

\author{Rui Xu$^{1,2}$}
 \email{rui.xu@ucl.ac.uk}
 \author{Sven Bestmann$^3$}
\author{Bradley E. Treeby$^1$}%
\author{Eleanor Martin$^{1,2}$}
\affiliation{$^1$Department of Medical Physics and Biomedical Engineering, University College London, London, UK}%
\affiliation{$^2$Wellcome/EPSRC Centre for Interventional and Surgical Sciences, University College London, London, UK}
\affiliation{$^3$Department of Clinical and Movement Neuroscience, University College London, London, UK}

\date{\today}

\begin{abstract}
Focused ultrasound spinal cord neuromodulation studies have demonstrated the capacity for neuromodulation of the spinal cord in small animals.
The safe and efficacious translation of these approaches to human scale requires an understanding of ultrasound propagation and heat deposition within the human spine. 
To address this, combined acoustic and thermal modelling was used to assess the pressure and heat distributions produced by a 500\,kHz source focused to the C5/C6 level of the cervical spine via two approaches a) the posterior acoustic window between vertebral posterior arches, or b) the lateral intervertebral foramen from which the C6 spinal nerve exits.
Pulse trains of 150 0.1\,s pulses with a pulse repetition frequency of 0.33\,Hz and free-field spatial peak pulse-averaged intensity of 10\,W/cm$^2$ were simulated for the CT volumes of four subjects and for $\pm$10\,mm translational and $\pm$10$^{\circ}$ rotational source positioning errors.
Target pressures ranged between 20\% and 70\% of free-field spatial peak pressures with the posterior approach, and 20\% and 100\% with the lateral approach.
When the source was optimally positioned with the posterior approach, peak spine heating values were below 1$^{\circ}$C, but source mis-positioning resulted in bone heating up to 4$^{\circ}$C.
Heating with the lateral approach did not exceed 2$^{\circ}$C within the mispositioning range.
There were substantial inter-subject differences in target pressures and peak heating values. 
Target pressure varied three to four-fold between subjects, depending on approach, while peak heating varied approximately two-fold between subjects.
This results in a nearly ten-fold range in the target pressure achieved per degree of maximum heating between subjects.  
This study highlights the importance of developing trans-spine ultrasound simulation software for the assurance of subject-specific safety and efficacy of focused ultrasound spinal cord therapies.
\end{abstract}

\maketitle


\section{Introduction}

Ultrasound can modulate neuronal behaviour\cite{takagi1960actions,ballantine1960progress,shealy1962reversible,colucci2009focused,tyler2008remote} and can be non-invasively focused to millimetre-sized volumes.
The small focal size that can be achieved with focused ultrasound is an advantage over non-invasive electric and magnetic neuromodulation techniques.
To achieve similar spatial precision with electric leads requires an invasive implantation procedure\cite{rowald2022activity}.
Electrical stimulation of the spinal cord has long been studied for movement restoration\cite{cybulski1984lower,rowald2022activity} and pain suppression\cite{tesfaye1996electrical,isagulyan2023effectiveness}.
Stimulation of the nearby dorsal root ganglia now competes with spinal cord stimulation\cite{deer2017dorsal,rowald2022activity,mekhail2021cost}, and has the advantage of the selectivity of the dorsal root ganglia versus the entire spinal cord. 
Focused ultrasound spinal cord and dorsal root ganglion neuromodulation may be a non-invasive and targeted alternative to existing approaches.
Pre-clinical studies demonstrate that ultrasonic spinal cord neuromodulation is possible\cite{takagi1960actions,ballantine1960progress,shealy1962reversible,liao2021effects,liao2021lifu,kim2022trans,tseha2023low,song2023transspinal}.
For example, it has been demonstrated that ultrasound can generate transient modulation of the descending tract in mouse spinal cords\cite{kim2022trans}, which has exciting implications for the treatment of movement disorders including Parkinson's disease and essential tremor.
Demonstrating the safety and efficacy of focused ultrasound spinal cord neuromodulation may enable future spinal cord therapies and studies on spinal cord function.

Low intensity, non-invasive ultrasonic spinal cord neuromodulation has not yet been performed in humans in peer-reviewed and published experiments.
Given the risks associated with the sensitivity of the spinal cord and the potential for heating due to the complex bony anatomy of the spine, it is crucial that a thorough safety assessment be performed prior to $in$ $vivo$ experiments. 
Many pre-clinical studies have investigated the bioeffects of ultrasound focused to the spinal cord, and the knowledge developed from these studies may provide a starting point for developing safety standards for the spinal cord\cite{xu2024safety}. 
Our recent review has shown that the threshold for reported ultrasound-induced damage in the spinal cords of a variety of animals, and a range of frequencies and environmental conditions can be approximated by an exponential equation of spatial-peak time-averaged ultrasound intensity and exposure time\cite{xu2024safety}. 
However, the human spine is larger and denser than the spines of animals used in the pre-clinical studies, and may generate ultrasound field aberrations that change the safety profile of potential ultrasonic neuromodulation approaches.

The human spine consists of stacked and highly irregular vertebrae that have contrasting acoustic properties from the surrounding soft tissue, causing the aberration of ultrasound wavefronts passing through to the spinal cord\cite{xu2018simulating}. 
In much of the spine, focusing ultrasound from the posterior approach is the only viable option due to the presence of ribs, lungs, and vertebral bodies\cite{xu2019spine}. 
Ultrasound can be focused through thoracic vertebral laminae using single-element focused transducers\cite{fletcher2018analysis}, but the foci tend to be shifted by several millimetres and occasionally split into sub-foci\cite{xu2018simulating}.
This resulted in inconsistent blood-spinal cord barrier opening in a study performed in the similar-to-human $in$ $vivo$ porcine spine that investigated the potential for improving therapeutic agent delivery to the spinal cord\cite{fletcher2020porcine}.
The success of similar targeted neuromodulatory applications in the thoracic spinal cord (e.g., modulation of sensation\cite{liao2021lifu} or motor control\cite{liao2021effects,kim2022trans,tseha2023low,song2023transspinal}) will require aberration correction methods to counteract the abberative effect of the spine.

\begin{figure*}
    \centering
    \includegraphics[width = 0.8\textwidth]{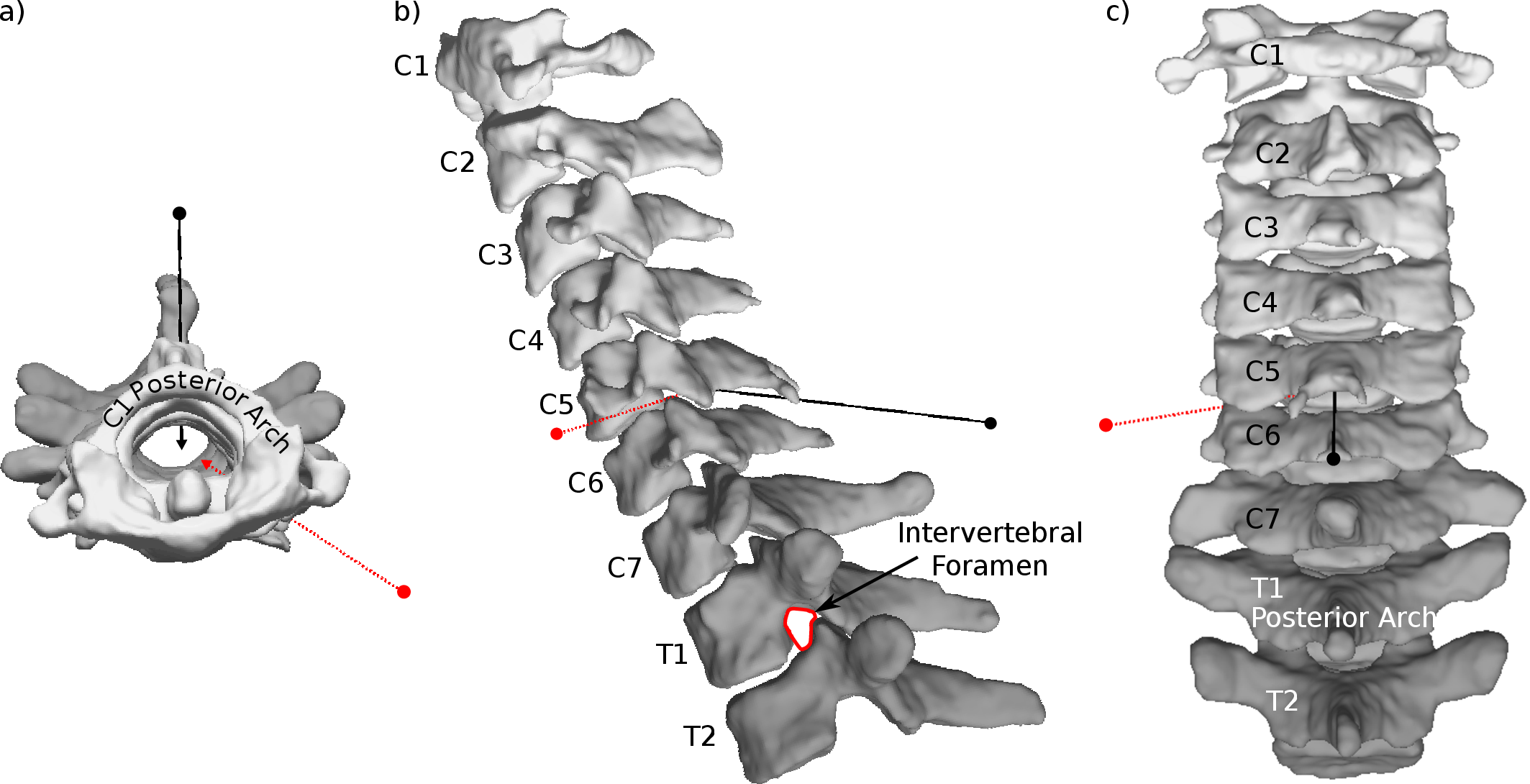}
    \caption{Potential lateral (red dotted line) and posterior (black solid line) ultrasonic approaches to the cervical spinal cord at the C5/C6 level, viewed from a a) superior, b) lateral, and c) posterior view. Mesh generated from sub\_verse651 (F, 37)\cite{liebl2021computed}.}
    \label{fig:TargetingPathsC5C6}
\end{figure*}

Ultrasonic approaches to lumbar spine have been investigated using simulation to estimate pressure distributions of ultrasound focused to intervertebral discs\cite{qiao2019delivering}.
Ultrasonic approaches to the human cervical spinal cord have not been addressed, and the morphology of the cervical spine offers several challenges and opportunities for ultrasound delivery.
The posterior elements of cervical vertebrae are angled relative to the neck skin surface and will complicate trans-spine transmission approaches that rely on normal wave incidence relative to the bone surface.
The air-filled trachea and esophagus preclude transmission from the anterior direction, and further transmission through the vertebral bodies or intervertebral discs will also be inefficient.
However, the cervical spine has posterior acoustic windows (between posterior arches) and lateral acoustic windows (through intervertebral foramen) that may enable the efficient delivery of ultrasound through the cervical spine to the spinal cord.
These acoustic windows are depicted at the C5/C6 level in Fig. \ref{fig:TargetingPathsC5C6}.

Here, we numerically assess the safety and efficacy of a low intensity ultrasonic pulse sequence focused to the human cervical spine at the C5/C6 level via both posterior and lateral approaches. 
We test a simple ultrasonic approach that does not implement spine aberration corrections and does not rely on optimized or stereotatic source positioning.
The safety and efficacy assessment is completed using combined acoustic and thermal simulation, and is based on subject-specific anatomy.

\section{Methods}

In this study, we develop a simulation framework for evaluating the safety and efficacy of ultrasound focused to potential cervical spinal cord neuromodulation targets, using acoustic and thermal metrics extracted from the simulations.
This simulation framework is based on \texttt{k-Plan} (Brainbox Ltd, Cardiff, UK), a graphical user interface-based ultrasound modelling tool marketed for transcranial ultrasound simulation, modified here for simulating ultrasound propagation through the spine. 
\texttt{k-Plan} implements k-Wave fluid simulations\cite{treeby2010k,treeby2012modeling}, and these simulations have been shown to perform accurately when compared to a multi-layered ray acoustics model for simulating trans-vertebral ultrasound propagation\cite{xu2018simulating,xu2022focusing}, and have been used in several simulation and experimental studies of ultrasound propagation through the spine\cite{xu2021array,xu2022focusing,frizado2023numerical}.
\texttt{k-Plan} also implements thermal simulations\cite{treeby2015contribution} which are suited to predicting pulse-induced heating.
The \texttt{k-Plan} simulations do not account for mode conversion and shear wave propagation within the spine.
Previous work has shown that the incorporation of shear waves did not substantially affect pressure values within the vertebral canal\cite{xu2022focusing}, but this work did not evaluate additional shear-wave induced heating within the spine. 
Elastic wave propagation was not simulated due to the high computational requirements needed for accurate elastic wave simulations; previous elastic transcranial simulations have used 25-60 spatial points per shear wavelength \cite{robertson2017sensitivity,jing2021effect}, while 8-10 points per longitudinal wavelength often suffices for transcranial simulation with the fluid k-Wave code\cite{robertson2017sensitivity,mcdannold2019elementwise,hosseini2023head}.

There are many parameters that can be chosen or optimized in a safety and efficacy simulation study.
The parameter list includes source geometry, source position, sonication frequency, sonication pulse parameters, sonication exposure duration, sonication intensity, not to mention the choice of target itself.
This study takes a treatment trial-like approach and fixes these parameters with $a$ $priori$ and $a$ $posteriori$ knowledge.
This approach does not generate a full investigation of the parameter space for focused ultrasound spinal cord therapies, but describes an approach that may be replicated with different parameters in future safety and efficacy studies.
This considered, the parameters chosen for this simulation study are intended to represent a sensible starting point for a focused ultrasound spinal cord neuromodulation study\cite{xu2024safety}.

The choice of source influences later choices on source positioning, source frequency, and targeting or aberration correction ability.
There is a trend towards developing application-specific sources.
For example, a simulation study optimized the design of a spine-specific phased array using simulations of ultrasound propagation through the thoracic spine\cite{xu2019spine}.
However, custom arrays can be time-consuming and expensive to develop and characterise and may not be needed for targets in the cervical spine where acoustic windows between vertebrae may simplify trans-spine ultrasound delivery.
Here, we simulate a commercially available 64-element array (H-313, Sonic Concepts, Seattle USA).
The H-313 array was packaged with the HIFUplex Plus 3000 system (Verasonics and Sonic Concepts, Seattle, USA).
The elements (13.34\,mm diameter) are arranged in an Archimedean spiral with an inner diameter of 44\,mm and outer diameter of 150\,mm, with a radius of curvature of 150\,mm.
It has a similar footprint to a spine-specific phased array optimized for the thoracic spine\cite{xu2019spine}, and the same operating frequency (500\,kHz). 

The \texttt{k-Plan} graphical user interface was used to position the source in simulation relative to four human spine x-ray computed tomography (CT) datasets obtained from the open access VerSe 2020 dataset\cite{liebl2021computed,loffler2020vertebral,sekuboyina2021verse}.
Two ultrasonic approaches were tested:
\begin{enumerate}
    \item a posterior approach: the source is positioned to focus sound through the acoustic windows between the C5 \& C6 posterior arches to the spinal cord
    \item a lateral approach: the source is positioned to focus sound through one of two intervertebral foramen to a C6 dorsal root ganglion
\end{enumerate}
The positioning of the source relative to the spines was performed visually in \texttt{k-Plan} by minimizing the intersection of element normal vectors with the spine.
Simulation may be used to optimise the placement of the source\cite{xu2019spine}, but here the objective was to evaluate safety with a source aligned by simpler means.
Trans-spine aberration correction\cite{xu2021array,frizado2023numerical} was not implemented here, but phasing was used in simulation to adjust the focal position of the source to the target assuming a free-field.
The experimental equivalent of this simulated approach requires the position of the source, the position of the target, and an estimated medium sound speed, which may be that of water or soft tissue.
Obtaining accurate experimental source and target positions relative to the subject anatomy is a separate challenge not addressed here.

A wide range of ultrasonic pulse regimes have been used to demonstrate a neuromodulatory effect in the spinal cords of small animal models\cite{xu2024safety}.
Here, 0.1\,s pulses are simulated, similar in time-scale to a pulse found to have both stimulative effects (in small pulse numbers) and inhibiting effects (in larger pulse numbers) in the seminal work performed at the Massachusetts General Hospital in the early 1960s\cite{ballantine1960progress,shealy1962reversible}.
The free-field spatial peak pulse-averaged intensity ($I_{\text{SPPA}}$) was set to 10\,W/cm$^2$ (548\,kPa), several-fold lower than used in the seminal work but in line with modern brain neuromodulation studies.
The pulse repetition interval was set to 3\,s, giving a pulse repetition frequency of 0.33\,Hz and a free-field spatial peak time-averaged intensity ($I_{\text{SPTA}}$) of 0.33\,W/cm$^2$.
This gives time for heat dissipation from hot spots that may form during each pulse.
Fifty pulses (150\,s total treatment time) were simulated, replicating an experimental design where a) averaging across pulses is needed in order to identify a neuromodulatory effect\cite{nandi2023ramped}, and/or b) the neuromodulatory effects increase throughout an exposure and an extended treatment time is needed\cite{tseha2023low}.
The combined total treatment time and free-field $I_{\text{SPTA}}$ was below the reported threshold for possible spinal cord damage in pre-clinical studies\cite{xu2024safety} (Supplementary Fig. \ref{fig:ispta_time_SIM}).

\subsection{Spine Simulation Medium Setup}

Spine computed tomography (CT) data and segmentations were obtained from an open-access repository intended for training automatic spine segmentation algorithms \cite{liebl2021computed}. 
A search through the repository was performed to identify CT datasets appropriate for trans-spine ultrasound simulation at the C5/C6 level.
Any dataset with a slice thickness over 1\,mm was excluded in order to maximize the accuracy of the voxelized representation of vertebral interfaces.
Any dataset that did not include the C5/C6 levels was also excluded.
The remaining datasets were sub\_verse549 (female, 48 years old, Siemens Somatom AS+, contrast enhanced - portal venous phase, 0.9766\,mm in-plane resolution, 0.6\,mm slice thickness), sub\_verse599 (male, 58 years old, Siemens Somatom AS+, contrast enhanced - portal venous phase, 0.8223\,mm in-plane resolution, 0.6\,mm slice thickness), sub\_verse618 (male, 28 years old, Philips ICT, contrast enhanced - portal venous phase, 0.9766\,mm in-plane resolution, 0.9\,mm slice thickness), and sub\_verse651 (female, 37 years old, Siemens Somatom AS+, not contrast enhanced, 0.5781\,mm in-plane resolution, 0.6\,mm slice thickness).
This sample contains two male CT scans (ages 28 and 58) and two female CT scans (ages 37 and 48). 
The chosen subjects cover a range of vertebral geometries and sizes, vertebral cortex thicknesses, and intervertebral disk heights.
This should increase the generalizability of the results in this study to a broader population. 

\begin{figure*}
    \centering
    \includegraphics[width = \textwidth]{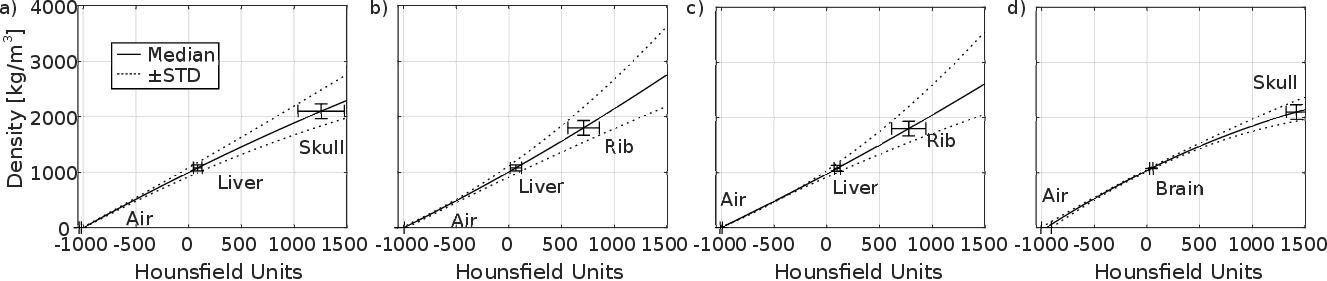}
    \caption{The mean and standard deviation (STD) in Hounsfield Units (HU) were obtained from air, liver, rib cortical bone, and skull cortical bone regions of interest for a) sub\_verse549 (F, 48), b) sub\_verse599 (M, 58), c) sub\_verse618 (M, 28), and d) sub\_verse651 (F, 37) in the VerSe 2020 dataset\cite{liebl2021computed}. The HU values are then plotted against mean literature density values and standard deviations\cite{ETB,hasgall2015database,wydra2013novel}.}
    \label{fig:SpineDensityCurves}
\end{figure*}

The spine acoustic properties were modelled heterogeneously.
Acoustic property maps were obtained from the CT scans using subject-specific conversion curves from Hounsfield Units (HU) to density.
The conversions curves were obtained from three regions of interest (ROI) from each subject (air, liver or brain, rib or skull cortical bone).
The soft tissue and cortical bone ROIs depended on the CT scanned volumes; if the skull and brain was present in the scan, they were used, otherwise rib and liver ROIs were used. 
The air, soft tissue, and cortical bone regions of interest were created in ITK-SNAP\cite{yushkevich2016itk}.
The density of air was assumed to be 1.2\,kg/m$^3$, the liver density was assumed to be 1079$\pm$53\,kg/m$^3$, the brain density was assumed to be 1046$\pm$6\,kg/m$^3$, the density of rib cortical bone was assumed to be 1800$\pm$133\,kg/m$^3$, and the density of the skull cortical bone was assumed to be 2100$\pm$133\,kg/m$^3$\cite{wydra2013novel,hasgall2015database}. 
The lower rib cortical density value was used to account for the thinness of the rib cortex relative to the skull cortex, resulting in greater partial volume effects within the rib ROIs than skull ROIs.
The HU to CT density calibration curves were generated with the Matlab \texttt{spline} function and are shown in Fig. \ref{fig:SpineDensityCurves}.
The standard deviation (STD) of the HU values in each subject is shown in Fig. \ref{fig:SpineDensityCurves}, along with the uncertainties in soft tissue and cortical bone density.
This ROI and subject-specific approach to obtaining a mapping from HU to acoustic properties was necessary to account for the use of different CT systems and reconstruction methods for the different subjects, and the absence of density calibrations within the VerSe dataset\cite{liebl2021computed}.

The VerSe 2020 CT datasets do not include soft tissue segmentations.
Soft tissue attenuates ultrasound and may contribute to field aberrations, so a homogeneous soft tissue layer was incorporated in each simulation.
The soft tissue segmentation for each subject was generated using semi-automatic segmentation in ITK-SNAP\cite{yushkevich2016itk}. 
A band-pass filter was first applied to the CT image (-200 to 100\,HU), then the soft tissues were manually seeded for the automatic contour evolution algorithm.
After semi-automatic segmentation, the masks were manually filled to ensure that the spines were fully encased in soft tissue. 
The soft tissue properties are listed in Table \ref{tbl:AcousticThermalProperties}.
Voxels outside the soft tissue masks were given the acoustic\cite{jones1992its,marczak1997water} and thermal\cite{huber2012new,kaye1926tables} properties of water.

The acoustic parameters (density, sound speed, and attenuation) of the spine were defined using the CT dataset.
The VerSe 2020 segmentation masks were binarized then element-wise multiplied by the CT volumes. 
The CT values were rounded to the nearest integer, binning the domain into 1\,HU steps.
The density, sound speed, and attenuation values were then applied in simulation to the binned 3D volumes to generate 3D maps of density, sound speed, and attenuation.
Density was mapped to the volumes using the calibration curves displayed in Fig. \ref{fig:SpineDensityCurves}.
Sound speed was then mapped to the volumes using a density ($\rho$) - sound speed ($c$) linear relationship optimized using a human spine\cite{xu2022establishing}; $c  = 0.35\rho + c_w$, where $c_w$ is the water sound speed.
A spine-specific attenuation function was not available, so a set of skull density-attenuation ($\alpha$) spline functions\cite{pichardo2010multi} were interpolated to 0.5\,MHz then mapped to the 3D volume. 

\begin{figure*}
    \centering
    \includegraphics[width = \textwidth]{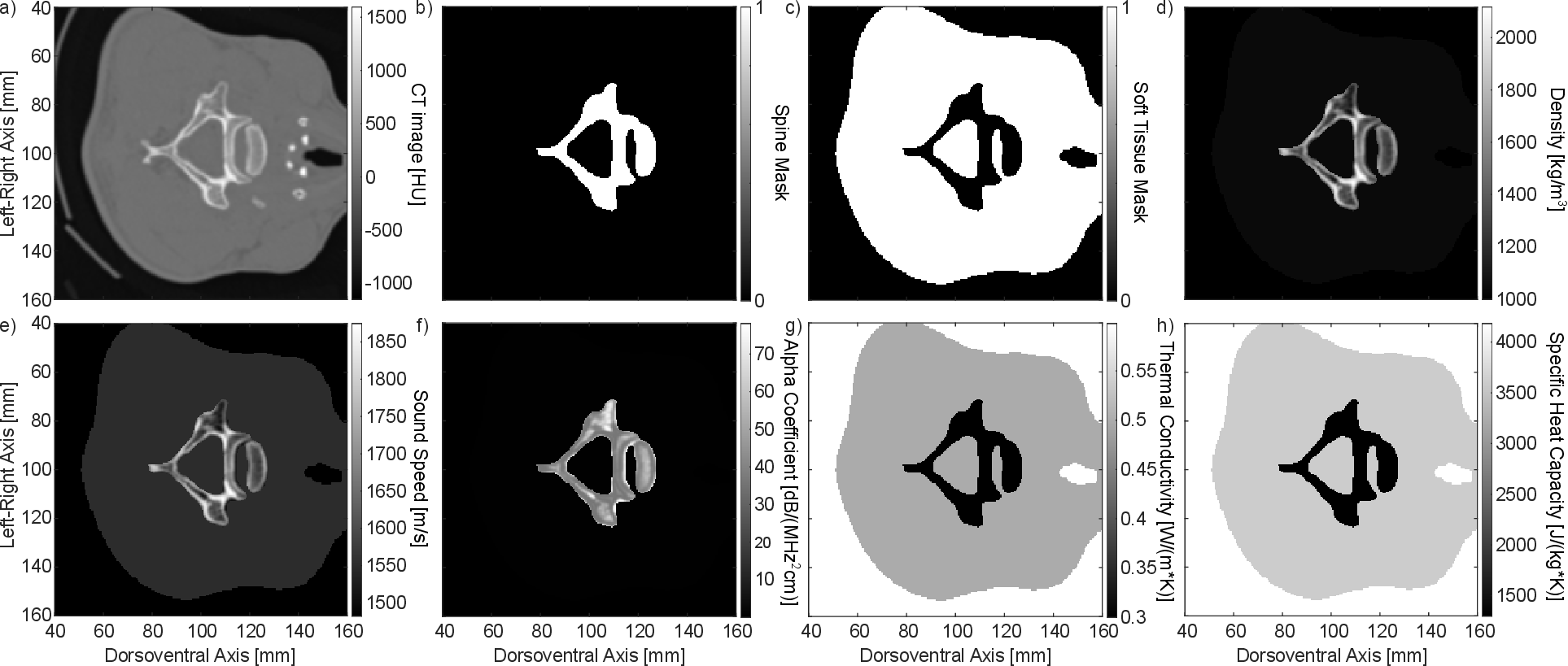}
    \caption{CT data from the VerSe 2020 dataset\cite{liebl2021computed} were used to generate the simulation domains. The a) CT image, b) spine segmentation, c) soft tissue segmentation, d) density (see Fig. \ref{fig:SpineDensityCurves}), e) sound speed\cite{xu2022establishing}, f) attenuation coefficient\cite{pichardo2010multi}, g) thermal conductivity and h) specific heat capacity\cite{hasgall2015database} of a transverse slice through the C5/C6 region of the sub\_verse651 (F, 37) spine.} 
    \label{fig:SpineSimulationSystem}
\end{figure*}

The spine thermal properties (density, thermal conductivity, and specific heat capacity) were defined homogeneously and the values are listed in Table \ref{tbl:AcousticThermalProperties}.
Both thermal conductivity and specific heat capacity were defined using the `worst-case scenario' for bone heating, i.e., the lowest specific heat value and thermal conductivity value found in the literature for bone\cite{hasgall2015database}.
Perfusion was not simulated, reducing the dissipation of heat away from any hot spots generated in the spine and soft spinal tissues.
The vertebral and spinal arteries and veins may act as substantial heat sinks, reducing local heating; this is not modelled.
It has been shown that ignoring perfusion resulted in a 4\% increase in focal heating in trans-skull heating simulations\cite{pulkkinen2014numerical}.
A slice from the CT image and the corresponding masks, acoustic property maps, and thermal property maps are shown in Fig. \ref{fig:SpineSimulationSystem}.

Surrounding bones (e.g. ribs, clavicle) are not included in the bone segmentation; their exclusion should not influence the simulated ultrasound and thermal fields at the C5/C6 level.
The H-313 array coupling cone was not included in the simulation domain and everything surrounding the soft tissue volume is modelled as water to avoid simulating ultrasound propagation at soft tissue-air boundaries.
This approximation does not have a large effect on the pressure fields as the coupling cone was designed to be wider than the source beam.
Air in the trachea and esophagus and surrounding the neck was modelled as water, again to avoid the fine spatial discretization needed to simulate boundaries with air. 
Neither posterior nor lateral approaches to the spinal cord intersect the esophagus or trachea and ultrasound energy will predominantly be absorbed by the spine.

\begin{table*}[ht]
\centering
\caption{Simulated acoustic and thermal properties of the spine, soft tissue, and water at 20$^{\circ}$C.}\label{tbl:AcousticThermalProperties}
\begin{tabular}{|l|c|c|c|}
\hline
Property                        &  Spine    & Soft Tissue & Water\\
\hline \hline
Density [kg/m$^3$]              & See text  & 1045\cite{hasgall2015database}        & 998\cite{jones1992its}     \\
Sound speed [m/s]               & See text\cite{xu2022establishing}  & 1550\cite{hasgall2015database}        & 1483\cite{marczak1997water}    \\
Absorption [db/cm.MHz$^y$]      & See text\cite{pichardo2010multi}  & 0.59\cite{hasgall2015database}        & 5.46e-4\cite{pinkerton1949absorption} \\
Absorption power law $y$        & 2\cite{treeby2010k}         & 1.2\cite{hasgall2015database}         & 2\cite{pinkerton1949absorption}       \\
Thermal conduction [W/m.K]      & 0.3\cite{hasgall2015database}       & 0.5\cite{duck1990physical}         & 0.598\cite{huber2012new}   \\
Specific heat [J/kg.K]          & 1300\cite{hasgall2015database}      & 3600\cite{duck1990physical}        & 4182\cite{kaye1926tables}    \\
\hline
\end{tabular}
\end{table*}

\subsection{Source Positioning}

\begin{figure*}
    \centering
    \includegraphics[width = \textwidth]{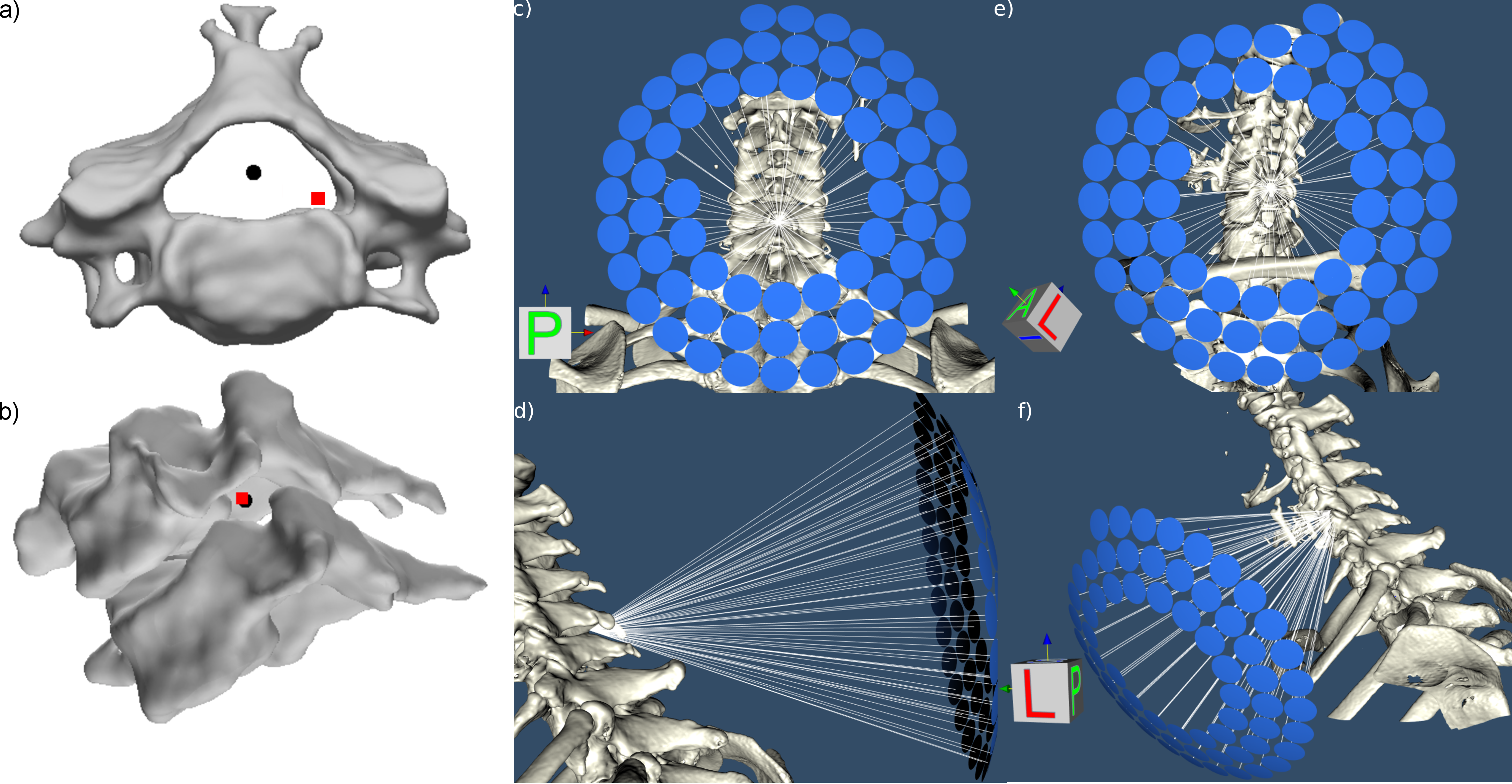}
    \caption{The 3D k-Plan transducer viewer and automatic mesh-generating software was used to place the transducer to target the spinal cord while minimizing geometric intersection of the element paths with the spine. a) Superior and b) lateral views of a posterior approach target (black circle) and lateral approach target (red square) at the C5/C6 level of the sub\_verse651 spine. c) and d) the source posterior approach, and e) and f) the lateral approach.}
    \label{fig:arraypositioning}
\end{figure*}

Targets were selected in \texttt{k-Plan} at the C5/C6 vertebral level of each spine (See Fig. \ref{fig:arraypositioning}a) and b)).
Central spinal cord targets were used when the source was positioned posterior to the spine, and anterio-lateral targets (approximating the location of a dorsal root ganglion within the intervertebral foramen) were used when the source was placed lateral to the spine.
The source was either a) oriented along the sagittal plane as shown in Fig. \ref{fig:arraypositioning}c,d) to focus ultrasound through the acoustic window between the C5 and C6 posterior arches, or b) oriented to focus ultrasound through the one of the C6 nerve intervertebral foramen, as shown in Fig. \ref{fig:arraypositioning}e,f).
For the posterior approach, the source was positioned by choosing a target in the spinal cord at the C5/C6 level, choosing a normal vector in the median plane and close to orthogonal to the skin surface that also minimized element normal intersection with the spine, then translating the source along the vector until the source was located at least 8\,cm (the height of the source coupling cone) outside of the body. 
This targeting procedure was intended to replicate a potential experimental method; the target vertebrae can be found using anatomical landmarks and the depth from skin to spinal cord can be estimated using ultrasonic imaging, MR or CT imaging, or using average depth values to the spinal cord at the C5/C6 level.
The lateral positioning was performed in a similar manner, placing the geometric focus of the source at the lateral target and minimizing the intersection of the element normal vectors with the spine.
These approaches assume that the coupling cone is sufficiently deformable to maintain contact with the skin surface at abnormal angles and for uneven skin surfaces.

\subsection{Simulation Parameters}

The \texttt{k-Plan} simulations used a spatial discretization of 6\,PPW (water sound speed = 1482.5\,m/s, giving a grid spacing 0.494\,mm in each dimension). 
Grid dimensions varied based on source orientation, but were at least 600$\times$600$\times$400 grid points to enclose the source and spine ROI.
\texttt{k-Plan} uses a default CFL number of 0.1 for acoustic and thermal simulations.
The acoustic simulations then use the k-Wave \texttt{checkStability} function to check simulation stability; and the temporal step is decreased if necessary.
The thermal simulations were stable with the CFL number of 0.1.
A convergence test was run from 4\,PPW, the minimum spatial discretization in \texttt{k-Plan}, to 9\,PPW, the maximum spatial discretization and total grid size enabled by the \texttt{k-Plan} off-site server for the given simulation domain sizes.

Four metrics were used to evaluate convergence of the pressure and thermal fields:
\begin{itemize}
    \item relative mean of the abs. difference between voxels: $\overline{|A_{x\text{PPW}} - A_{\text{9PPW}}|}/\text{max}(A_{\text{9PPW}})$
    \item relative L$_{\infty}$ norm:  max$|A_{x\text{PPW}} - A_{\text{9PPW}}|/\text{max}(A_{\text{9PPW}})$
    \item relative difference in target values:  $|A_{x\text{PPW}}^{tar} - A_{\text{9PPW}}^{tar}|/\text{max}(A_{\text{9PPW}}^{tar})$
    \item relative abs. peak difference: $|\text{max}(A_{x\text{PPW}}) - \text{max}(A_{\text{9PPW}})|/\text{max}(A_{\text{9PPW}})$
\end{itemize}
The pressure and thermal fields were re-sampled to a 0.5\,mm$^3$ isotropic grid for comparison.
The target values were obtained from a 3x3x3mm$^3$ (approximately 1 wavelength, cubed) volume rather than the target point to account for shifts in the pressure field resulting from shifts in the standing wave pattern generated within the vertebral canal, caused by the re-sampling of the mask representing the vertebral interface.

\begin{figure*}
    \centering
    \includegraphics[width = \textwidth]{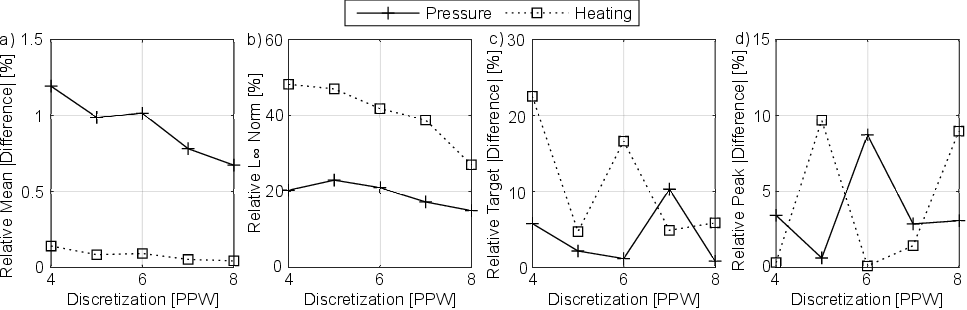}
    \caption{a) Relative mean absolute difference between simulation results at the given spatial sampling and simulations results at nine points per wavelength (PPW). b) L$_{\infty}$ norm, c) relative target absolute difference, and d) relative absolute peak difference.}
    \label{fig:convergence}
\end{figure*}

The target and maximum pressure and temperature values (Fig. \ref{fig:convergence}c,d) are used in later analyses within this work.
At 6\,PPW, these values differ by 0-15\% from the 9\,PPW values, and simulations at even 4\,PPW do not vary substantially from the 9\,PPW values.
The L$_{\infty}$ norm is the most stringent of the tested metrics; L$_{\infty}$ norm values (See Fig. \ref{fig:convergence}b) remain high within the tested discretization range due to the resampling of the spine mask at each new discretization, and the quasi-continuous nature of the simulated pulse and resulting standing wave formation within the vertebral canal.
Resampling of the mask changes the relative position of the bone interfaces, changing the position of interference patterns and standing waves that have spatial frequencies double that of the source frequency. 
Achieving an L$_{\infty}$ norm of 10\% with a simpler spherical scatterer geometry has been shown to require at least 10\,PPW\cite{treeby2020nonlinear}, beyond the computational resources available for this study.  
The choice was made to perform the simulations at 6\,PPW to reduce the computational burden of the thermal simulations while maintaining good thermal accuracy and pressure amplitude accuracy. 
At 6\,PPW, one acoustic simulation requires a total computation time, including pre- and post-processing steps, of at least two hours using an NVIDIA A40 (Ampere) 48\,GB GPU in an Intel Xeon CPU E5-2620 @ 2.40\,GHZ server.
At 6\,PPW, one thermal simulation requires a total computation time, including pre- and post-processing steps, of at least four hours using the aforementioned compute server.
Over 200 simulations were completed to generate the results in this study, necessitating reasonable computation times for individual simulations.

\subsection{Quantification of Pressure and Temperature Fields}

\texttt{k-Plan} simulations were used to assess the variations in pressure and heating that result from variations in source position, changes in sonication parameters, uncertainties in acoustic properties, and different spine morphologies.
The \texttt{k-Plan} simulations generate 3D steady-state pressure amplitude fields encapsulating the entire source and target simulation domain.
This information is condensed by extracting parameters of interest from the 3D pressure volumes.
The two primary pressure output parameters are the pressure at the target, which gives an estimate of spine-induced attenuative and aberrative losses, and the spatial peak pressure, which may also have safety implications for cavitation, particularly in the presence of gas-filled microbubbles. 
Additional pressure metrics were implemented for select simulations.
For the posterior approach where standing wave formation is significant, we used a metric to estimate the amplitude of standing waves at the focus.
The ratio of the standing wave magnitude to the DC component of the pressure amplitude distribution was obtained from the pressure profile along a 10\,mm vector centred at the target and aligned with the source.
The magnitude of the frequency component corresponding to the standing waves was calculated and divided by the mean pressure along the vector. 
The vector length was chosen to fit within the average C6 vertebral canal\cite{ulbrich2014normative}.
Supplementary Fig. \ref{fig:standingwaves} illustrates the extraction of the standing wave magnitude relative to the DC pressure amplitude at the focus.
For the posterior approach, we used a 3D cylindrical mask (5\,mm diameter\cite{ulbrich2014normative}, 10\,mm height) to isolate spatial peak pressure values and focal shifts within the spinal cord.
Standing waves and spinal cord volumes were not assessed for the lateral approach, as the geometry is less liable to the formation of standing waves and the target is located in the smaller dorsal root ganglion, rather than the spinal cord. 
Focal shifts for the lateral approach were defined as the distance between target and spatial peak pressure location.

Maximum temperature rises at the target and spatial peak temperature locations were extracted from the simulation domains, along with distances between targets and maximum heating locations.
An additional metric, termed `thermal efficiency' and defined as the target pressure achieved per degree of maximum heating was computed to calculate the efficacy of the ultrasonic approach. 

\subsubsection{Positioning Accuracy.}\label{sec:tolerance}

Simulations were performed to investigate the pressure and temperature field sensitivity to source position. 
For the posterior approach, the lateral source coordinates were fixed in alignment with the midline (no rotation around the craniocaudal axis) and the anteroposterior distance from source to target was fixed at the focal length.
Due to the spine geometry, errors are more likely to arise in vertical positioning, so simulations were performed with $\pm$10\,mm vertical source shifts in 2.5\,mm increments, and $\pm$10$^{\circ}$ rotational shifts around the frontal axis in 2.5$^{\circ}$ increments.
This generated 17 source positions per subject and spinal cord target. 
For the lateral approach, vertical source shifts ($\pm$10\,mm in 2.5\,mm increments) and rotational shifts around the vertical axis ($\pm$10$^{\circ}$ in 2.5$^{\circ}$ increments) were simulated.
This generates 17 source positions per subject and dorsal root ganglion target.
Free-field phase corrections were applied by the array elements to maintain the focus at the target.
Resulting differences in sonication efficiency and heating are thus due to changes in ultrasound path intersection with the spine.

\subsubsection{Uncertainty.}\label{sec:uncertainty}

Simulations were performed to investigate the pressure and temperature field sensitivity to uncertainty in the acoustic property maps. 
Uncertainties in the simulations primarily originate from the mapping from HU to density, then from density to sound speed\cite{xu2022establishing} and from density to attenuation\cite{pichardo2010multi}.
Computational methods have been developed to assess the effect of simulation uncertainties on the output acoustic (and potentially thermal) fields\cite{robertson2017sensitivity,stanziola2023transcranial}.
Here, HU uncertainty propagation was performed in a similar way to Robertson $et$ $al.$ 2017\cite{robertson2017sensitivity}.
Variations in the HU-to-density functions were defined by the standard deviations in values displayed in Fig. \ref{fig:SpineDensityCurves} for each subject.
HU-density splines were fit to ($\overline{\text{HU}} + 1$\,STD, $\overline{\rho} - 1$\,STD) of each ROI to create a `minimum spline', and to ($\overline{\text{HU}} - 1$\,STD, $\overline{\rho} + 1$\,STD) of each ROI to create a `maximum spline' for each subject (dashed lines in Fig. \ref{fig:SpineDensityCurves}).
Simulations at the central (no translation, no rotation) source locations were repeated with the maximum and minimum splines.
The updated $\rho$ values are coupled to the sound speed and attenuation functions\cite{xu2022establishing,pichardo2010multi}, generating new sound speed and attenuation maps.
This approach does not account for additional uncertainties in the sound speed and attenuation conversions from density, or from uncertainties in the thermal parameters.
It tests only the extremes of the HU-$\rho$ functions within the $\pm$1 STD ROI range.
Fully sampling the HU-$\rho$ distribution requires a Monte Carlo approach\cite{stanziola2023transcranial} or linear uncertainty propagation with a differentiable simulator\cite{stanziola2023transcranial,stanziola2023j}, which was beyond the scope of this work.

\subsubsection{Intensity.}

Simulations were performed to quantify changes in the temperature rise at different $I_{\text{SPPA}}$ values.
Simulations at the central (no translation, no rotation) source locations were repeated at free-field $I_{\text{SPPA}}$ values of 5, 10, and 20\,W/cm$^2$, which correspond to free-field spatial peak pressure amplitudes of 387, 548, and 775\,kPa.
The \texttt{k-Plan} simulations do not simulate non-linear ultrasound propagation, so pressure fields were re-scaled linearly to obtain the different free-field peak pressure amplitudes.
Heat deposition ($Q)$ via ultrasonic absorption is modelled using the plane wave assumption $Q = \alpha p^2 / (\rho_0c_0)$ where $\alpha$ is the attenuation coefficient in Np/m, $\rho_0$ is the mass density, $c_0$ is the sound speed, and $p$ is the pressure amplitude.

\section{Results}

The spatial peak pressure, pressure at the target, and spatial peak temperature are averaged across all positions for each approach and for each subject and are displayed in Table \ref{tbl:SimResults_ArrayShifts} and Figure \ref{fig:boxplots_posterior_lateral}.

\begin{table}[ht]
\centering
\caption{Spatial peak pressure ($p_{\text{max}}$), target pressure ($p_{\text{tar}}$), and maximum heating ($\Delta T_{\text{max}}$) for the 10\,W/cm$^2$ free-field spatial peak pulse-averaged intensity (548\,kPa) simulations averaged over 17 posterior (post.) or 17 lateral (lat.) positions. Mean values $\pm$ one standard deviation.  }\label{tbl:SimResults_ArrayShifts}
\footnotesize
\begin{tabular}{|l|c|c|c|c|}
\hline
Metric                             &  F, 48    & M, 58 & M, 28 & F, 37 \\
\hline \hline
$p_{\text{max}}$ (post.) [kPA]  &  402$\pm$29 & 372$\pm$31 & 305$\pm$21 & 548$\pm$55 \\ 
$p_{\text{tar}}$ (post.) [kPA]  &  113$\pm$24 & 115$\pm$13 & 141$\pm$26 & 317$\pm$27 \\
$\Delta T_{\text{max}}$ (post.) [$^{\circ}$C]  &  1.78$\pm$0.88 & 0.81$\pm$0.09 & 1.45$\pm$0.58 & 1.87$\pm$0.51 \\    
$\Delta T_{\text{tar}}$ (post.) [$^{\circ}$C]  &  0.10$\pm$0.04 & 0.07$\pm$0.05 & 0.11$\pm$0.02 & 0.28$\pm$0.05 \\    
$p_{\text{max}}$ (lat.) [kPa]   &  373$\pm$48 & 334$\pm$20 & 497$\pm$55 & 549$\pm$25 \\  
$p_{\text{tar}}$ (lat.) [kPa]   &  277$\pm$35 & 106$\pm$24 & 251$\pm$23 & 428$\pm$63 \\ 
$\Delta T_{\text{max}}$ (lat.) [$^{\circ}$C]   &  0.43$\pm$0.07 & 1.37$\pm$0.11 & 1.50$\pm$0.18 & 0.65$\pm$0.12 \\       
$\Delta T_{\text{tar}}$ (lat.) [$^{\circ}$C]   &  0.10$\pm$0.01 & 0.12$\pm$0.02 & 0.12$\pm$0.01 & 0.20$\pm$0.01 \\       
\hline
\end{tabular}
\end{table}

\subsection{Posterior Approach}

\begin{figure*}
    \centering
    \includegraphics[width = \textwidth]{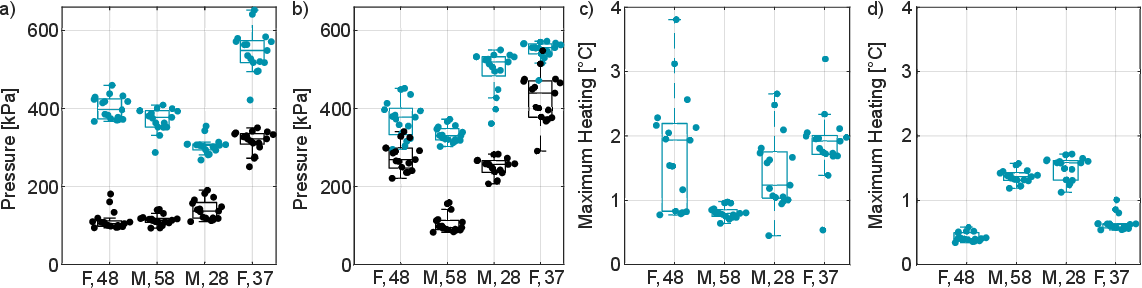}
    \caption{Maximum (blue) and target (black) pressures and maximum heating for each source position and subject, for the a,c) posterior and b,d) lateral source positions. Boxplots represent the median and 25$^{\text{th}}$/75$^{\text{th}}$ quartiles.}
    \label{fig:boxplots_posterior_lateral}
\end{figure*}

\begin{figure*}[ht]
     \centering
    \includegraphics[width = \textwidth]{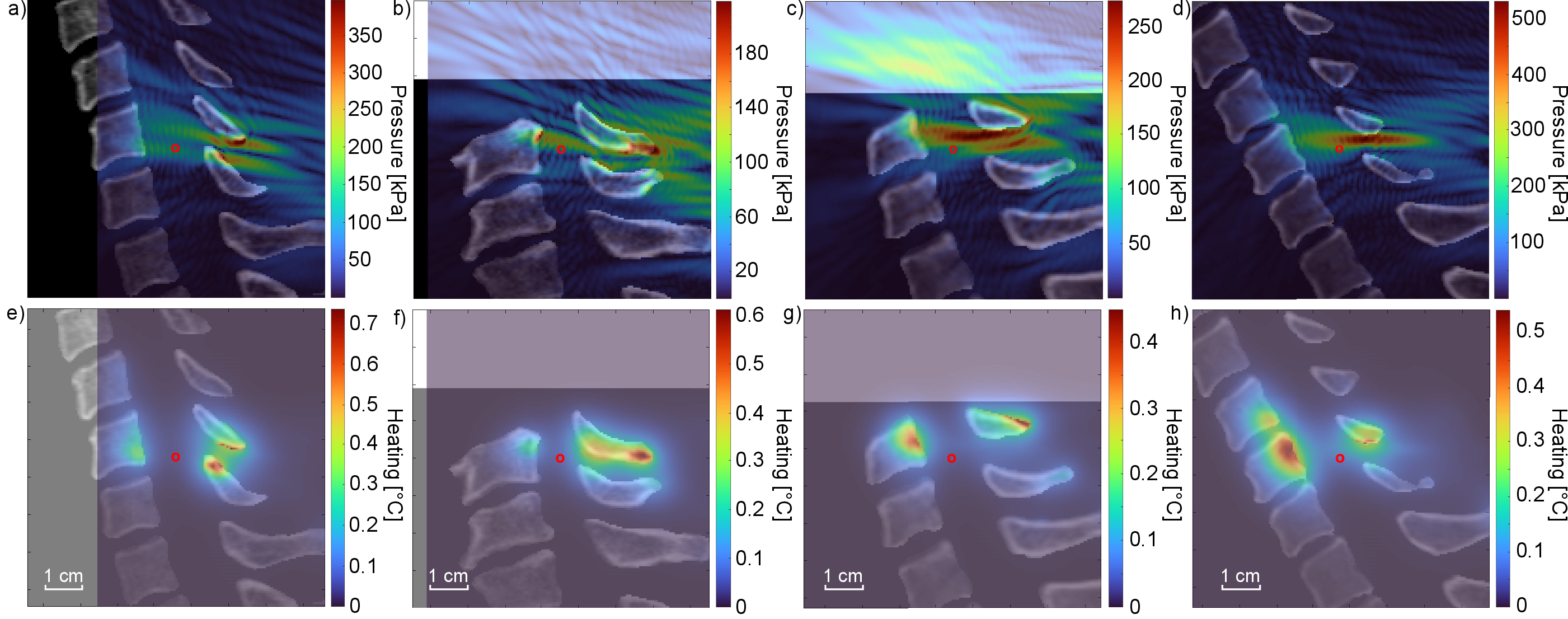}
    \caption{Posterior source positions. Lateral views of the pressure amplitude and heating spatial distributions overlaid over the masked CT scans of a,e) sub\_verse549 (F, 48), b,f) sub\_verse599 (M, 58), c,g) sub\_verse618 (M,28), and d,g) sub\_verse651 (F, 37). }
    \label{fig:Posterior_pressure_heat}
\end{figure*}

The posterior approach gives spinal geometry-dependent access to the spinal cord.
Fig. \ref{fig:boxplots_posterior_lateral}a) and c) display the spatial peak pressure amplitude, the pressure amplitude at the target, and the spatial peak temperature for each subject.
The $in$ $situ$ target pressure amplitudes were approximately 20\% of the free-field spatial peak pressure in three subjects (F, 48; M, 58; M, 28).
This transmission value is lower than the 32\% reported experimentally for individual thoracic vertebrae and a single bowl transducer focused through the posterior arch\cite{xu2018simulating}.
This substantial transmission loss suggests a limited efficacy of the posterior approach in these three subjects. 
However, in one subject (sub\_verse651, F, 37), the mean $in$ $situ$ target pressure amplitude was 317\,kPa, giving a much higher mean transmission efficiency of 58\%. 
There is a visible difference in the spine geometry in this subject; a larger and more favourable acoustic window through the spine to the spinal cord (Fig. \ref{fig:Posterior_pressure_heat}).
It may be possible to improve transmission to the spinal cord in the three low-efficacy subjects with appropriate neck flexion to increase the size of the posterior acoustic windows\cite{xu2019spine}.

\begin{figure*}[ht]
    \centering
    \includegraphics[width = \textwidth]{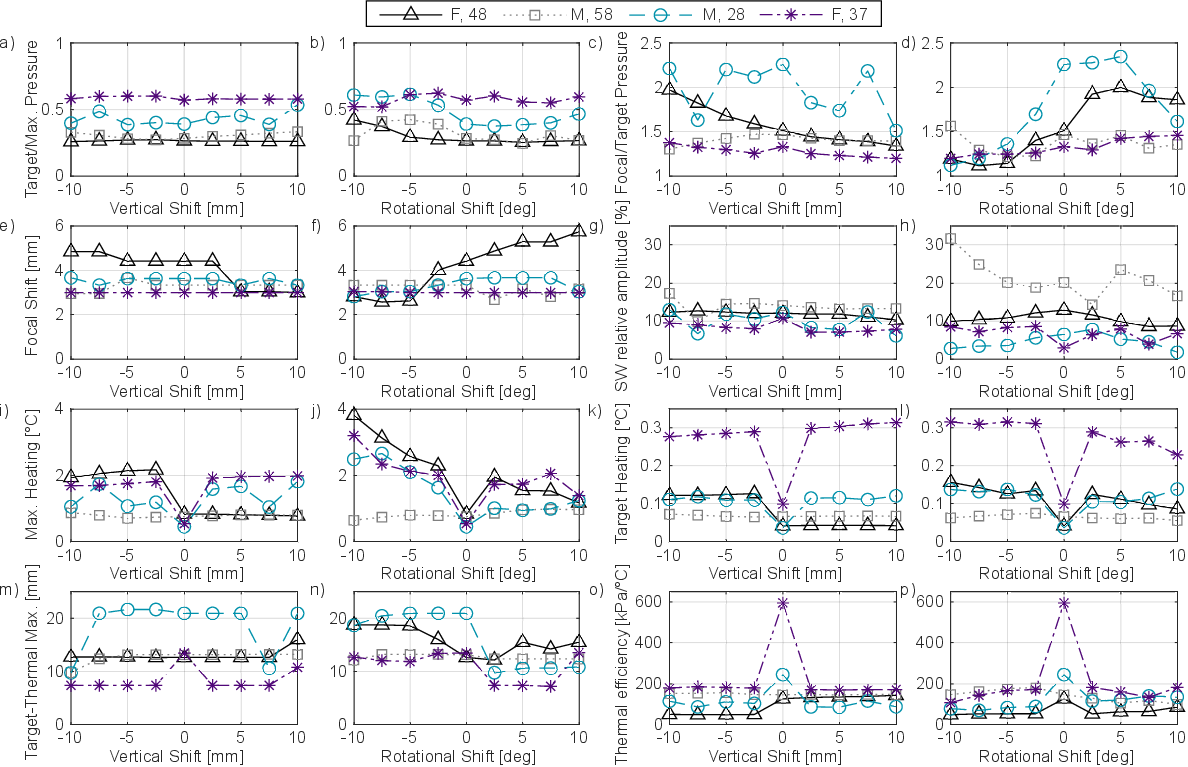}
    \caption{Pressure and heating metrics for the tested posterior source positions. a,b) the ratio of target pressure to spatial peak pressure, c,d) the ratio of spinal cord focus pressure to target pressure, e,f) the focal shift within the spinal cord relative to the target position, g,h) the standing wave (SW) amplitude relative to the DC pressure along the source axis within the spinal cord, i,j) the maximum heating within the simulation domain, k,l) the target heating, m,n) the distance between target and peak heating location, and o,p) the pressure at the target per degree of maximum heating.}
    \label{fig:PostArray_Position_Results}
\end{figure*}

Figure \ref{fig:PostArray_Position_Results} shows the effect of rotation ($\pm$10$^{\circ}$ in 2.5$^{\circ}$ increments) and vertical translation ($\pm$10\,mm in 2.5\,mm increments) on four pressure metrics and four thermal metrics.
Source vertical translation and rotation had little influence on the ratio of target pressure to spatial peak pressure.
This may be because waves from each element entering the waveguide-like acoustic windows still arrive at the target relatively in-phase.
Lateral slices through the target positions in the pressure and temperature distributions are shown in Fig. \ref{fig:Posterior_pressure_heat}.
The spatial peak pressure location tended to be outside the spinal cord, but focal pressures within the spinal cord (Fig. \ref{fig:PostArray_Position_Results}c,d) also reached up to nearly 2.5 times the pressure at the target.
The spine-induced pressure field aberrations all demonstrate focal shifts in the superior direction.
This may be due to the angling of the spinous processes and posterior arch interface. 
The focal shifts within the spinal cord ranged between 2\,mm and 6\,mm (Fig. \ref{fig:PostArray_Position_Results}e,f).
Therefore, spinal cord bioeffects are likely to occur within a few millimetres of the intended target locations.
Causes of focal shifts are standing wave formation within the vertebral canal (Fig. \ref{fig:PostArray_Position_Results}g,h), which mostly contributed 10-15\% of the DC pressure amplitude within the vertebral canal at the target (but up to over 30\% in one subject), as well as aberration and attenuation resulting from trans-spinal ultrasound transmission.
The variability in standing wave relative amplitude and the other pressure metrics between subjects and between source positions illustrates the utility of acoustic simulations in subject-specific treatment planning.

Maximum heating values range similarly between all subjects, with the exception of one subject (sub\_verse599, M, 58), where the maximum heating was consistently around 1$^{\circ}$C.
Heat tends to be deposited in both the posterior arch and the vertebral bodies.
When sound is more efficiently delivered to the vertebral canal, heat deposition in the vertebral bodies increases.
This is most obvious in sub\_verse651 (F, 37).
Heating in the surrounding soft tissues appears to mostly result from heat dissipation from nearby bone.

The maximum heating across all source positions and subjects is 4$^{\circ}$C.
If we were to conservatively assume heating of 4$^{\circ}$C for the entire 150\,s pulse train (and a base temperature of 37$^{\circ}$C), the maximum cumulative equivalent minutes (CEM) at 43$^{\circ}$C\cite{sapareto1984thermal} in bone would be 0.16 CEM43$^{\circ}$C, lower than the suggested brain neuromodulation soft tissue threshold of 0.25 CEM43$^{\circ}$C\cite{aubry2023itrusst}.
Soft tissue CEM43$^{\circ}$C values are further below the suggested brain neuromodulation CEM43$^{\circ}$C threshold\cite{aubry2023itrusst}.

The heating at the target and in the cylindrical spinal cord volume is always substantially lower than the peak heating values (See Fig. \ref{fig:PostArray_Position_Results}) and follows the same source-position trends as the peak heating values, albeit with different maximum-to-target heating ratios.
The correlation coefficient (calculated with \texttt{corrcoef}) for the maximum and target heating values was calculated to see if more heating in bone results in more target heating.
The correlation coefficient calculated across all maximum and target heating values is 0.62; when separating the heating values by subject, the median of the correlation coefficients for maximum and target heating values is 0.83.
This indicates that spinal cord heating predominantly originates from heat absorption in bone which then dissipates into the soft tissues; this effect was previously seen in pre-clinical spinal cord ablation studies\cite{fry1953temperature,borrelli1986ultrasonically}.
The distance between target and maximum heating location (Fig. \ref{fig:PostArray_Position_Results}m,n) ranges between 8\,mm and over 20\,mm (all outside the spinal cord cylinder), further supporting the finding that peak heating occurs in bone.
Target heating is highest and target-thermal maximum distance is lowest in sub\_verse651 (F, 37), where the maximum heating location is on the posterior surface of the vertebral body.
Maximum heating locations in the other subjects were on the posterior surface of the vertebral arches. 
The variability in the thermal metrics between subjects and between source positions illustrates the utility of combined acoustic and thermal simulations in subject-specific treatment planning.

There is substantial inter-subject variability in target pressure and heating. 
This is emphasized in Fig. \ref{fig:PostArray_Position_Results}o,p) which depicts the target pressure achieved per degree of maximum heating in the simulation domain.
Efficiency is visibly improved at the central source positions that minimise array element normal intersection with the spine.
These results demonstrate that the posterior approach is sensitive to source position, particularly when the spine is in a neutral position.
Neck flexion may increase the size of the acoustic windows, reducing the inter-subject variability in target pressure and heating and reducing the sensitivity of the approach to source position, but this remains to be demonstrated with CT images of flexed necks.

\subsection{Lateral Approach}

\begin{figure*}[ht]
    \centering
    \includegraphics[width =\textwidth]{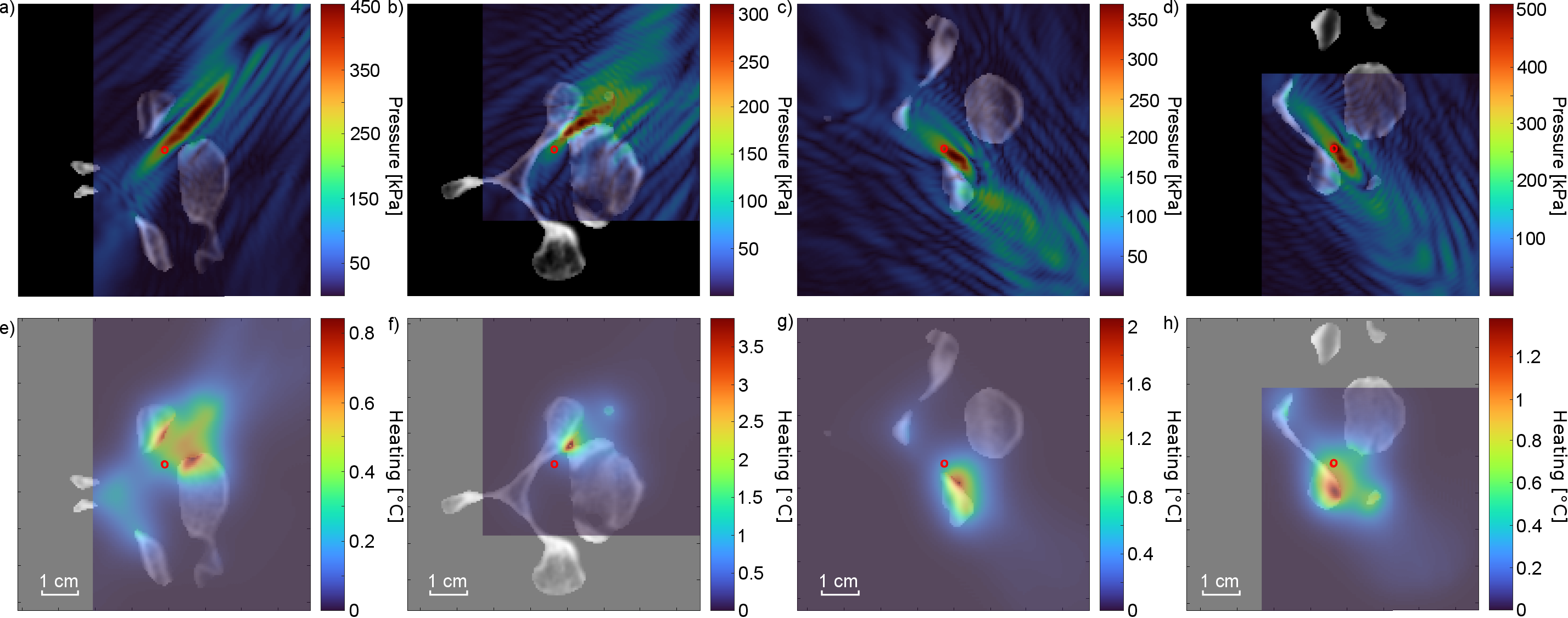}
    \caption{Lateral source positions. Superior views of the pressure amplitude and heating spatial distributions overlaid over the masked CT scans of a,e) sub\_verse549 (F, 48), b,f) sub\_verse599 (M, 58), c,g) sub\_verse618 (M,28), and d,g) sub\_verse651 (F, 37).}
    \label{fig:Lateral_pressure_heat}
\end{figure*}

Seventeen acoustic and thermal simulations were performed for each subject with the lateral approach.
The mean and standard deviation in spatial peak and target pressure and temperature are reported in Table \ref{tbl:SimResults_ArrayShifts}, and in Fig. \ref{fig:boxplots_posterior_lateral}.
Figure \ref{fig:Lateral_pressure_heat} displays the pressure and heat distributions on 2D planes through the targets for the `central' positions of the tested translation and rotation ranges for the four subjects.

Figure \ref{fig:Lateral_pressure_heat}a-d) show that the dorsal root ganglion is relatively accessible via the lateral approach, and the pressure distributions generally display less aberration than those shown in Fig. \ref{fig:Posterior_pressure_heat} for the posterior approach.
The focus intersects the target in three of four cases; the intervertebral foramen in the oldest of the four subjects (sub\_verse599; M, 58) appears smaller and the focal region does not reach the target.
Figures \ref{fig:Lateral_pressure_heat}a-d) also show that standing waves do not visibly influence the pressure distribution at the focus, likely due to the oblique incidence on the posterior arch of the C6 vertebra.

The dorsal root ganglion target is quite close to bone, and consequently the heat deposition field is centred around the target with clear heat dissipation from the bone to the soft tissue target area. 
However, because the spatial peak pressure locations are in soft tissue rather than bone, the maximum heating values are lower on average than for the posterior approach (See Fig. \ref{fig:boxplots_posterior_lateral}c,d and Table \ref{tbl:SimResults_ArrayShifts}).

\begin{figure*}[ht]
    \centering
    \includegraphics[width =\textwidth]{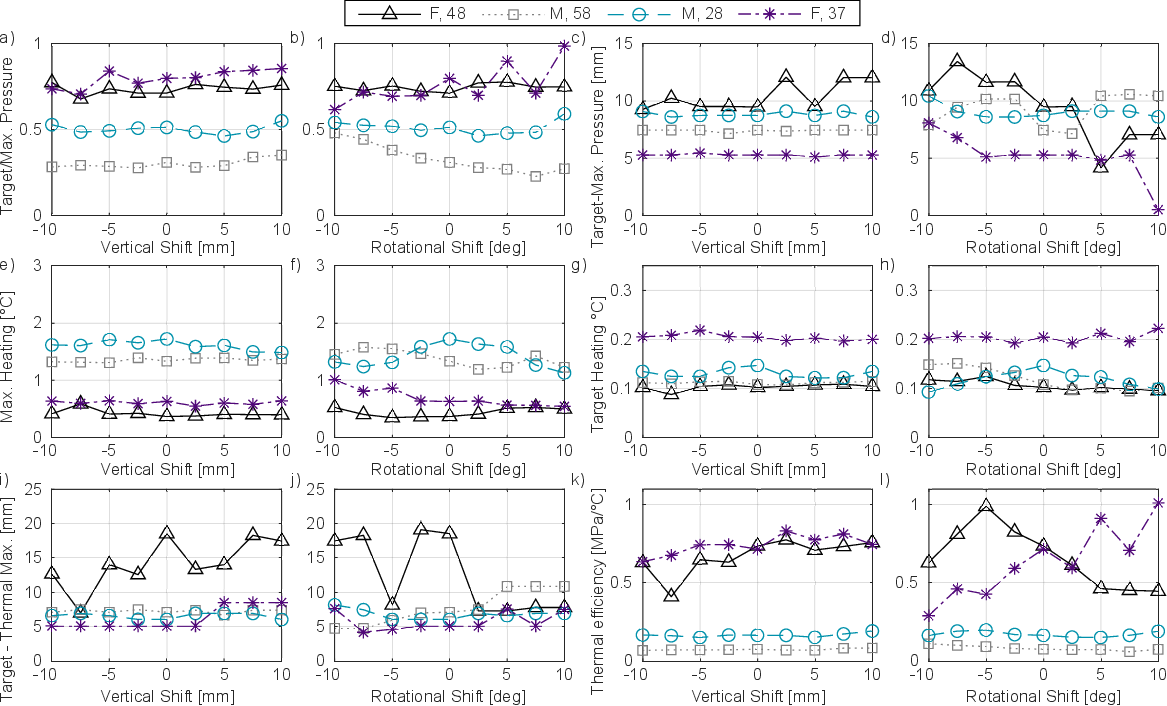}
    \caption{Pressure and heating metrics for the tested lateral source positions. a,b) the ratio of target pressure to spatial peak pressure, c,d) the distance between target and spatial peak pressure location e,f) the maximum heating within the simulation domain, g,h) the target heating, i,j) the distance between target and peak heating location, and k,l) the pressure achieved at the target per degree of maximum heating.}
    \label{fig:LatArray_Position_Results}
\end{figure*}

Figure \ref{fig:LatArray_Position_Results} shows the effect of source alignment on pressure and thermal metrics.
None of the identified metrics trend consistently with source alignment across all subjects, although there is some variation within the subjects.
This differs from the posterior source position, which was considerably more thermally efficient in the identified source positions that minimized element normal intersection with the vertebrae. 
There is still considerable inter-subject variability, with target pressures and heating values varying several-fold between subjects.

\subsection{Uncertainty}

Simulations were repeated at the central source positions for each subject, but with mappings from HU to density defined instead with their maximum splines ($\overline{\text{HU}} + 1$\,STD, $\overline{\rho} - 1$\,STD) and minimum splines ($\overline{\text{HU}} - 1$\,STD, $\overline{\rho} + 1$\,STD).

\begin{figure*}
    \centering
    \includegraphics[width = \textwidth]{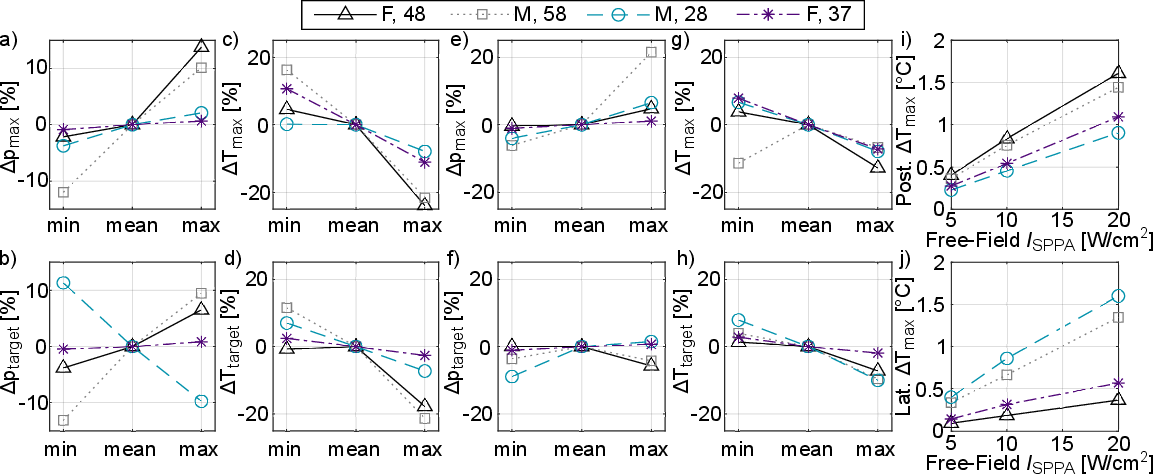}
    \caption{Uncertainty estimation and heat scaling with intensity. The changes in spatial peak pressure ($p_{\text{max}}$), target pressure ($p_{\text{target}}$), peak heating (T$_{\text{max}}$), and target heating (T$_{\text{target}}$), are displayed for the posterior approach (a-d) and the lateral approach (e-h). The relationships between free-field spatial-peak pulse-averaged intensity and T$_{\text{max}}$ are displayed in i) and j) for the posterior and lateral approaches.}
    \label{fig:uncertainty_intensity}
\end{figure*}

Spatial peak pressures rise as the spline HU-$\rho$ conversions shift towards a higher density for a given HU, for both approaches (Fig. \ref{fig:uncertainty_intensity}a,e).
This effect likely results from an increase in impedance mismatch at the soft tissue-bone interfaces, resulting in greater reflections from bone interfaces and more ultrasound transmission through the acoustic windows to the vertebral canal.
The range in spatial peak pressure is approximately $\pm$20\% for the $\pm$1\,STD range in conversion spline. 
Figure \ref{fig:uncertainty_intensity}b,f) does not show a trend in target pressure with conversion spline, but does show a range of variation of approximately $\pm$10\% for the $\pm$1STD range in conversion spline. 

Conversely, peak heating decreases as the spline HU-$\rho$ conversions shift towards a higher density for a given HU (Fig. \ref{fig:uncertainty_intensity}c,g).
This effect also likely results from the increase in impedance mismatch at the soft tissue-bone interfaces, resulting in less ultrasound transmission into bone, where the ultrasound is quickly attenuated and converted to heat.
The range peak heating is approximately $\pm$20\% for the $\pm$1\,STD range in conversion spline.
Target heating continues to correlate with peak heating values (See Fig. \ref{fig:uncertainty_intensity}d,h), and the trends seen in peak heating persist in target heating despite there being no direct connection between changes in HU-$\rho$ conversion, other than the heat dissipation from bone hot spots.
These results also show that the simulations are not very sensitive to variations in acoustic properties.

\subsection{Intensity}

Figure \ref{fig:uncertainty_intensity}i,j) shows the maximum heating for 5, 10, and 20\,W/cm$^2$ free-field $I_{\text{SPPA}}$.
The peak heating values scale linearly with intensity.
\texttt{k-Plan} does not simulate non-linear ultrasound propagation. 
This assumption is valid when the pressure amplitude is substantially less than $\rho_0 c_0^2$\cite{cobbold2006foundations}, which in water means substantially less than 1.5\,MPa.
The pressure amplitudes corresponding to the simulated $I_{\text{SPPA}}$ values of 5, 10, and 20\,W/cm$^2$ in water are 387, 548, and 775\,kPa.
For the source dimensions, a 15\,cm free-field path length, and a spatial peak pressure amplitude of 775\,kPa, over 95\% of the wave energy will remain at the source fundamental frequency, simulated with the k-Wave axisymmetric code\cite{treeby2020nonlinear}.
At higher intensities and if the peak pressure location coincides with bone, energy in higher harmonics will be attenuated at greater rates and result in greater but more localized heating. 
Ignoring non-linear ultrasound propagation may generate substantial differences in peak thermal dose in high intensity applications\cite{pinton2011effects}, but is less consequential at the simulated pressures in this study.

\section{Discussion}

Simulations of spinal cord neuromodulation pulse trains focused to the C5/C6 level were completed for four subjects to evaluate the resulting pressure and heat distributions.
Target pressures ranged between 20\% and 70\% of free-field spatial peak pressure amplitudes with the posterior approach, and 20\% and 100\% with the lateral approach.
Trans-skull transmission values at approximately 0.5\,MHz have been measured at 36-48\%\cite{white2006longitudinal,riis2022acoustic,chen2023numerical} and analytically estimated to peak at approximately 60\%\cite{attali2023three}.
The trans-spine transmission values range from lower than trans-skull transmission, to slightly higher with the posterior approach, and to substantially higher with the lateral approach.  
The posterior approach can be effective if the source is positioned to focus through the intervertebral gaps and avoid bone, but this still depends on spine geometry and acoustic window size.
The spatial peak pressure was shifted from the target location, often outside the spinal cord, even when using the posterior acoustic window.
Ultrasonic spinal cord neuromodulation with a simple source and without aberration correction will be limited by acoustic and thermal safety in the surrounding tissues. 
The limits in the surrounding tissues partially depend on the hypothetical risk versus reward of a pulse sequence; some skull heating\cite{connor2004patterns,hughes2018reduction} is acceptable in ultrasonic thalamotomies due to the benefit derived from the procedure. 
Ultrasonic spinal heating above thresholds developed for ultrasonic brain neuromodulation will also only be acceptable with meaningful subject benefit. 

We tested ranges of source positions to model a source aligned by imperfect means. 
Pressures at spinal cord targets tend not to vary strongly with source position, but maximum heating values increase rapidly as the source position is shifted from the central heuristically optimal positions. 
Heating at the spinal cord targets remains minimal.
However, some spinal cord targets are inaccessible when focusing through the intervertebral gap, and the posterior arches of cervical vertebrae are not well suited to transvertebral ultrasound transmission.
This work shows that the posterior approach may be implemented for certain human C5/C6 targets without necessitating spine aberration corrections.
This may facilitate future spinal cord neuromodulation studies, in both small animals and potentially in humans.

This work also investigated a lateral approach and found that it is less sensitive to source position than the posterior approach, but still relies on the accurate identification of the target position relative to the source elements.
The target pressures are higher with lateral approach than the posterior approach, on average, and the lateral approach is more thermally efficient, on average. 
The ultrasonic neuromodulation of the dorsal root ganglion appears to be viable at human scale\cite{hellman2021pilot}, which may have exciting implications for targeted neuromodulation of motor control\cite{rowald2022activity} and sensation\cite{hellman2021pilot}.

Research on the combined acoustic and thermal simulation of ultrasound propagation through the spine to the spinal cord is limited.
Fletcher $et$ $al$. used the k-Wave elastic code\cite{treeby2014modelling,treeby2015contribution} \texttt{pstdElastic2D} to simulated ultrasound propagation in 2D through a porcine thoracic vertebra lamina, estimating heating of up to 0.33$^{\circ}$C in the porcine vertebral canal for their implemented blood-spinal cord barrier opening pulses.
Several other works investigated ultrasound propagation through the human spine\cite{xu2018simulating,qiao2019delivering,adams2020silico,xu2022establishing,frizado2023numerical} but did not investigate heating.
There is, however, a substantial body of work on combined acoustic and thermal simulation of ultrasound propagation through the human cranium\cite{connor2004patterns,pernot2004prediction,pinton2011effects,pulkkinen2011simulations,pulkkinen2014numerical,kyriakou2015full,ding2015modulation,mueller2016computational,hughes2018reduction,mcdannold2019elementwise,hosseini2023head}.
Several of these studies investigate high-intensity ultrasound applications (e.g., ultrasonic thalomotomy to treat essential tremor)\cite{connor2004patterns,pulkkinen2014numerical}, with intensities that may necessitate non-linear modelling\cite{pinton2011effects} and temperature-dependent acoustic and/or thermal properties\cite{hughes2018reduction,mcdannold2019elementwise}.
These additional complexities should not be necessary for low-intensity spinal cord neuromodulation sequences, but may be applied in future spinal work if necessary.
There are fewer studies on the combined acoustic and thermal modelling for ultrasonic brain neuromodulation\cite{mueller2016computational,hosseini2023head}, perhaps due to the lower implemented intensities and thermal risks, resulting in a greater focus instead on the acoustic field aspects of the sonications.
Useful simulation tools have been developed for brain neuromodulation, including conservative pressure transmission estimates through a template skull averaged from 20 humans skulls for different beam sizes and source frequencies between 0.1-1.5\,MHz\cite{attali2023three}, and a head template from 29 different skulls for both acoustic and thermal simulation\cite{hosseini2023head}
Representative spine templates could be useful in the development of safety and efficacy guidelines for trans-spine sonications; there is much to learn from the trans-skull literature.

There are several limitations to the simulation approach implemented in this work.
Uncertainties in acoustic properties were partially addressed by simulating different conversion curves for HU to density.
This approach does not address additional uncertainties in the density to sound speed conversion function\cite{xu2022establishing}, which have been established with a combined simulation-experiment approach using ultrasound propagation through the posterior arches of one set of $ex$ $vivo$ human thoracic vertebrae.
Additional uncertainties will arise in the density to attenuation conversion function\cite{pichardo2010multi}.
This is based on skull attenuation rather than spine attenuation, as spine-specific conversions curves have not yet been developed.
Uncertainties in thermal properties were not modelled, but conservative values (low thermal conductivity, low specific heat values) that amplify heating effects were used throughout the simulation study.
Efficient uncertainty estimation\cite{stanziola2023transcranial} may be a useful tool in the further development of simulation as a tool for trans-spine ultrasound treatment planning and safety standard development.

Elastic wave propagation was not modelled due to computational constraints.
Elastic wave modelling has been shown to generate small differences in trans-vertebral pressure distributions\cite{xu2022focusing} and shear waves were found to contribute an average of 11\% of the velocity magnitude transmission to the thoracic vertebral canal via the vertebral laminae\cite{xu2022establishing}.
With the acoustic window approach studied here, shear waves are less likely to deliver sound to the vertebral canal and more likely to exacerbate spinal heating due to the higher critical angle for mode conversion to shear waves in the spine\cite{treeby2015contribution}, and higher attenuation rate of shear waves\cite{pichardo2017viscoelastic}.
Future experiments that employ temperature field monitoring may help to untangle the relative effects of the simulated (conservative) thermal properties and the exclusion of shear wave propagation.

This simulation study presents a safety analysis of two ultrasonic approaches to the cervical spinal cord with a single 500\,kHz source and sonication scheme. 
There are an indefinite number of permutations of spinal targets, spinal geometries, source positions, source frequencies, and source geometries, along with sonication sequences. 
The applicability of our results to new studies will vary with deviation from the parameters simulated here. 
For example, higher frequencies and smaller focal sizes may fit better within the cervical spine acoustic windows, resulting in less heating, but could also generate greater bone heating if the foci are poorly positioned.
The pulse train used here had a relatively long inter-pulse interval that gives heat ample time to dissipate from hot spots forming within the spine.
Sonications with similar $I_{\text{SPTA}}$ values may generate similar final heat distributions, but the CEM43$^{\circ}$C will vary. 
This simulation study included four subjects, which is unlikely to represent the full range of anatomy across sexes and ages, as suggested by the substantial inter-subject differences in pressure and heating in Fig. \ref{fig:boxplots_posterior_lateral}.
Future ultrasonic spinal cord therapies will likely require subject and sonication-specific simulation to ensure the efficiency and safety of the approach for a therapy that employs intensities and exposure durations near the reported threshold for possible damage\cite{xu2024safety}.
These simulations will likely need to account for uncertainties in source position  (simulated here), and may also need to account for changes in spine geometry from the spine geometry captured in the images used to generate the spine simulation domains (not simulated here).
We hope that this study provides an outline for a simulation approach to evaluate the safety and efficacy of a focused ultrasound spinal cord application. 

\section{Summary}

The safe and efficacious translation of pre-clinical focused ultrasound spinal cord therapies to human scale requires an understanding of ultrasound propagation and heat deposition within the human spine. 
We developed a combined acoustic and thermal simulation approach to assess the pressure and heat distributions produced by a 500\,kHz source focused to the C5/C6 level of the cervical spine via a) the posterior acoustic window between vertebral posterior arches, or b) the lateral intervertebral foramen from which the C6 spinal nerve exits.
Potential neuromodulatory pulse trains consisting of 150 0.1\,s pulses with a pulse repetition frequency of 0.33\,Hz and free-field $I_{\text{SPPA}}$ of 10\,W/cm$^2$ were simulated for four subjects.
Both approaches are intended to reduce aberration and attenuation and to simplify the hardware and beamforming requirements for focused ultrasound spinal cord neuromodulation. 
Target pressures ranged between 20\% and 70\% of free-field spatial peak pressure amplitudes with the posterior approach, and 20\% and 100\% with the lateral approach.
When the source was optimally positioned with the posterior approach, peak heating values within the spine were below 1$^{\circ}$C, but source translation and rotation resulted in heating up to 4$^{\circ}$C.
Heating with the lateral approach did not exceed 2$^{\circ}$C for a $\pm$10\,mm translational and $\pm$10$^{\circ}$ rotational range.
There were substantial inter-subject differences in target pressures and peak heating values, highlighting the importance of the development of subject-specific trans-spine ultrasound simulation software for the assurance of the safety and efficacy of potential focused ultrasound spinal cord therapies.

\section{Acknowledgements}

This work was supported in part by a UKRI Future Leaders Fellowship [grant number MR/T019166/1], in part by the Wellcome/EPSRC Centre for Interventional and Surgical Sciences (WEISS) (203145Z/16/Z), and in part by the EPSRC, UK. 
This work was also supported by the Ministry of Education, Youth and Sports of the Czech Republic through the e-INFRA CZ (ID:90254).
For the purpose of open access, the author has applied a CC BY public copyright licence to any Author Accepted Manuscript version arising from this submission.
The authors wish to thank Dr. Carys Evans for useful discussions on this work. 



\bibliography{references}

\appendix

\section{Supplementary Figures}
\setcounter{figure}{0}    

Supplementary figures are included here in lieu of a separate supplementary data file. 
\begin{figure*}[ht]
    \centering
    \includegraphics[width = 0.7\textwidth]{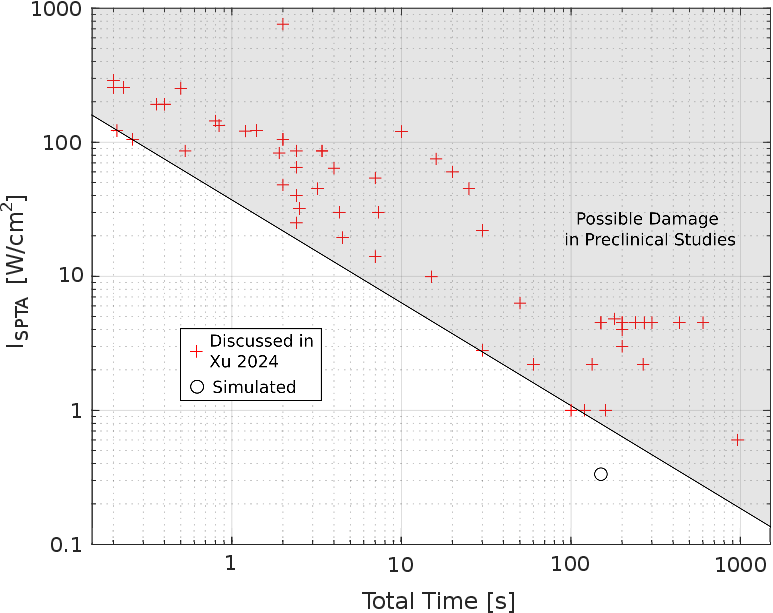}
    \caption{The simulated free-field spatial peak time averaged intensity (I$_{\text{SPTA}}$) and total treatment time was below the reported threshold for damage in preclinical studies, discussed in: Xu, Rui, Bradley E. Treeby, and Eleanor Martin. "Safety review of therapeutic ultrasound for spinal cord neuromodulation and blood–spinal cord barrier opening." Ultrasound in Medicine \& Biology (2024)\cite{xu2024safety}.}
    \label{fig:ispta_time_SIM}
\end{figure*}

\begin{figure*}
    \centering
    \includegraphics[width = 0.8\textwidth]{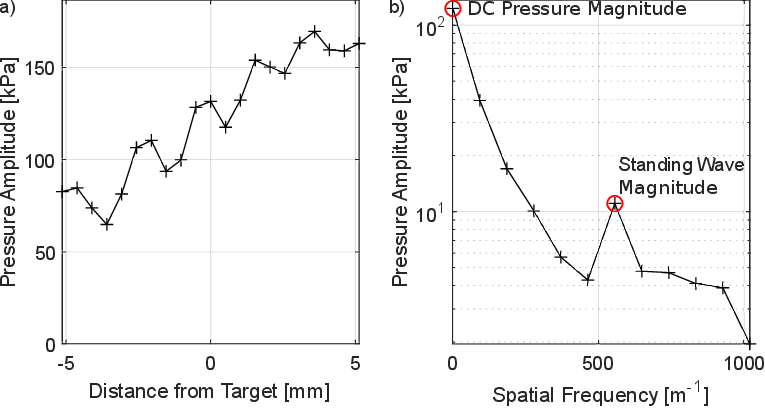}
    \caption{Standing waves are generated from reflections from the vertebral bodies. The ratio of standing wave amplitude to DC pressure amplitude through the focus is extracted from a 10 mm vector aligned with the source, centred at the intended focus. The standing wave has a spatial frequency double that of the source. This figure depicts one example of standing wave extraction.}
    \label{fig:standingwaves}
\end{figure*}

\end{document}